\documentclass[11pt,a4paper]{article}
\input{epsf.tex}
\usepackage{latexsym}
\newtheorem{thm}{Theorem}
\begin{document}

\title{The Edmonds asymptotic formulae for the $3j$ and $6j$ symbols}
\author{James P. M. Flude\thanks{email. james.flude@maths.nott.ac.uk}\\Department of Mathematics\\University of Nottingham\\University Park\\Nottingham\\NG7 2RD, UK}

\maketitle

\begin{abstract}
The purpose of this paper is to provide definitions for, and proofs of, the asymptotic formulae given by Edmonds, which relate the $3j$ and $6j$ symbols to rotation matrices.
\end{abstract}

\newpage

\section{Introduction}

The purpose of this paper is to provide definitions for, and proofs of, the
asymptotic formulae given by Edmonds \cite{edm55}, which relate the $3j$ and $6j$ symbols to rotation matrices.

The study of the asymptotics of the $6j$ symbol dates back to Racah \cite{racah51},
who stated an asymptotic formula relating $6j$ symbols and Legendre
polynomials. Edmonds \cite{edm55} later generalized Racah's formula by replacing
the Legendre polynomial by a rotation matrix. For $a,b,c\gg f,m,n$, Edmonds
stated, 
\begin{equation} \label{1}
\left\{ 
\begin{array}{ccc}
c & b & a \\ 
f & a+n & b+m
\end{array}
\right\} \approx \frac{\left( -1\right) ^{a+b+c+f+m}}{\sqrt{\left(
2a+1\right) \left( 2b+1\right) }}d_{nm}^f\left( \phi \right)
\end{equation}
where 
\begin{equation} \label{2}
\cos \phi =\frac{a\left( a+1\right) +b(b+1)-c(c+1)}{2\sqrt{a(a+1)b(b+1)}}.
\end{equation}
Edmonds also gave a further result which related the $3j$ symbol to a
rotation matrix. For $a,c\gg b$, Edmonds stated, 
\begin{equation} \label{3}
\left( 
\begin{array}{ccc}
b & a & c \\ 
\beta & \alpha & \gamma
\end{array}
\right) \approx \frac{\left( -1\right) ^{b-a-\gamma }}{\sqrt{2c+1}}d_{\delta
,-\beta }^b\left( \theta \right)
\end{equation}
where 
\begin{equation} \label{4}
\cos \theta =\frac{ \gamma}{\sqrt{c(c+1)}}.
\end{equation}
Edmonds did not give a proof of either result, nor did he define how the
angular momenta scaled or how the $3j$\thinspace and $6j$ symbols approached
the rotation matrices.

Edmonds reasoned that, for sufficiently large $c$, (\ref{4}) could be replaced
by 
\begin{equation} \label{5}
\cos \theta =\frac{ \gamma}{ c}.
\end{equation}
Brussaard and Tolhoek \cite{bruss}, following a suggestion of Edmonds, argued
heuristically, that if $a,c\gg b$ and $\theta $ is defined by (\ref{5}) then (\ref{3}) holds. Brussaard and Tolhoek did not define how $a$ and $c$ scaled, nor did they define how the $3j$ symbol approached the rotation
matrix.

Some years later, Ponzano and Regge \cite{pr} discussed the geometry of the
asymptotics of the $3j$ and $6j$ symbols. To each asymptotic $6j$ symbol,
Ponzano and Regge associated a tetrahedron whose edges had length $j+\frac 12
$, where $j$ is an entry in the $6j$ symbol, and to each asymptotic $3j$
symbol they associated a triangle whose edges were labelled in the same
manner. Thus, the tetrahedron associated to the above $6j$ symbol is that given in figure 1, from which

\begin{equation} \label{6}
\cos \phi =\frac{\left( a+n+\frac{1}{2}\right) ^2+\left( b+m+\frac{1}{2}\right)
^2-\left( c+\frac{1}{2}\right) ^2}{2\left( a+n+\frac{1}{2}\right) \left( b+m+\frac{1}{2}\right) }.
\end{equation}

\begin{figure}[htb!]
\hspace{3cm} \epsfxsize=8cm \epsfbox{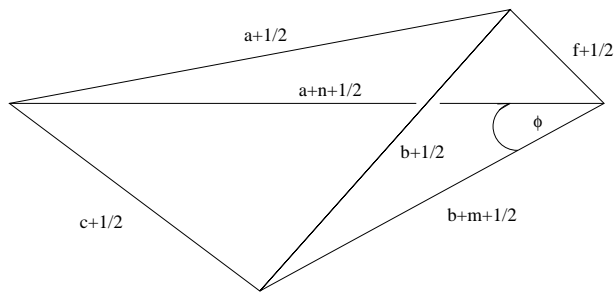}
\caption{\label{fig1} Associated tetrahedron}
\end{figure}

The triangle associated to the above $3j$ symbol is given in figure 2, from which

\begin{equation} \label{7}
\cos \theta =\frac{\gamma}{c+\frac{1}{2}}.
\end{equation}

\begin{figure}[htb!]
\hspace{5cm} \epsfysize=6cm \epsfbox{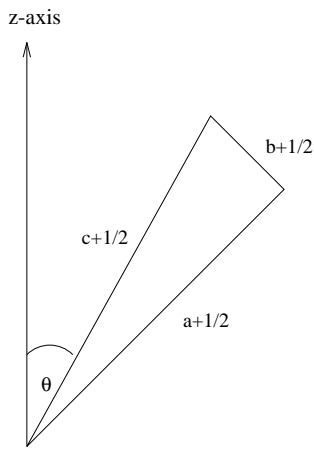}
\caption{\label{fig2} Associated triangle}
\end{figure}

Hence there is a discrepancy between the geometrical aspect, given by
Ponzano and Regge, and the algebraic aspect, given by Edmonds, Brussaard and
Tolhoek. If we relabel the edges of the associated tetrahedron and triangle
by $\sqrt{j(j+1)}$ instead of $j+\frac{1}{2}$ then $\phi $ and $\theta $ are
defined by (\ref{2}) and (\ref{4}) respectively. In [$4$],
Ponzano and Regge give an important asymptotic formula for the $6j$ symbol
in which the associated tetrahedron is label as above. If we relabel the
edges of the associated tetrahedron by $\sqrt{j(j+1)}$ and apply the
relevant modifications to Ponzano and Regge's asymptotic formula, then this
modified formula provides a worse approximation to the $6j$ symbol than
Ponzano and Regge's original formula. This is discussed in the first part of
section four. We define $\phi $ and $\theta $ by (\ref{6}) and (\ref{7}) respectively, rather than by (\ref{2}) and (\ref{4}) for the following two reasons. The first is that all the formulae are simpler with these definitions and the second is that these definitions are consistent with Ponzano and Regge which is important for the above reasons.

The purpose of this paper is to provide definitions for, and proofs of, the
asymptotic formulae which relate the $3j$ symbol and rotation matrix of (\ref{3}), and the $6j$ symbol and rotation matrix of (\ref{1}). In both cases, we define how the angular momenta scale, and we also define how these particular $3j$ and $6j$ symbols approach the rotation matrices of (\ref{3}) and (\ref{1}) respectively. These definitions make precise $a,c\gg b$ and $a,b,c\gg f,m,n$ and $\approx $. With these definitions, in section $2$, we prove (\ref{3}), where $\theta $ is defined by (\ref{7}), and in section $3$, we prove (\ref{1}), where $\phi $ is defined by (\ref{6}). In the first part of section $4$, we discuss why $\phi $ and $\theta $ are defined by (\ref{6}) and (\ref{7}) rather than by (\ref{2}) and (\ref{4}). In the last part of section $4$, we discuss the relationship between the results proved in sections $2$ and $3$. This entails using the result proved in section $2$ to arrive at an expression of a similar type to that proved in section $3$.

The study of the asymptotics of $3j$ and $6j$ symbols is important in
understanding the Ponzano-Regge model of three dimensional Euclidean quantum
gravity [$4$]. In this approach, three dimensional spacetime is approximated
by a simplicial manifold formed by gluing together tetrahedra. To each edge
of a tetrahedron in this simplicial manifold Ponzano and Regge associate a
half integer $j$, with the length of that edge defined as $j+\frac 12$, in
such a way that the triangle inequality is satisfied on all the faces of
that tetrahedron. They then thus associate to each such labeled tetrahedron
a $6j$ symbol.

Any potential candidate for the quantum theory of gravity must reduce to
general relativity in some suitable limit. By allowing various combinations
of the edges in each tetrahedron in the simplicial manifold to become large,
the Ponzano-Regge model reduces to three dimensional general relativity
without a cosmological constant. Thus, to understand the relationship
between the Ponzano-Regge model of quantum gravity and general relativity
the asymptotics of the $3j$ and $6j$ symbols is important. The reduction of
the above asymptotic $6j$ symbol to the rotation matrix $d_{nm}^f(\theta )$
plays a crucial role in understanding the relationship between the
Ponzano-Regge model of three dimensional quantum gravity and the
construction of topological state sums from tensor categories \cite{jwbc97}.

Figure $3$ \cite{jwbfig} shows all the possible asymptotic limits of the $6j$ symbol. These limits correspond to moving various combinations of the four vertices
of the associated tetrahedron. At the top of figure $3$ is the unscaled $6j$
symbol in which the four vertices remain fixed. In the $1+3$ case, one
vertex is taken to infinity while the remaining three remain fixed. This
represents the reduction of a $6j$ symbol to a $3j$ symbol [$4$]. In the $%
1+1+2$ case, two of the vertices remain fixed and the remaining two move
away from the fixed vertices as well as away from each other. This
represents (\ref{1}).

\newpage

\begin{figure}[ht!]
\hspace{3cm} \epsfysize=7cm \epsfbox{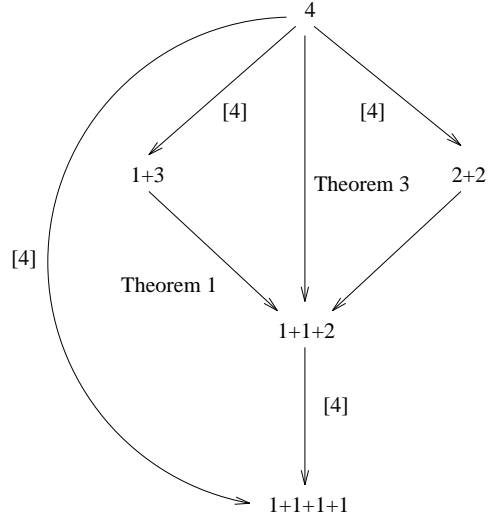}
\caption{\label{fig3} Asymptotics of the $6j$ symbol}
\end{figure}

\section{Rotation matrices and $3j$ symbols}

Following Edmonds [$1$], we first define the $3j$ symbol and then the
rotation matrix.

Let $a,b$ and $c$ be non-negative half integers which satisfy the triangle
inequality, and let $\left| \alpha \right| \leq a,\left| \beta \right| \leq
b $ and $\left| \gamma \right| \leq c$ with $\alpha +\beta +\gamma =0$, then
the $3j$ symbol is defined by 
\begin{eqnarray} \label{8}
\left( 
\begin{array}{ccc}
a & b & c \\ 
\alpha & \beta & \gamma
\end{array}
\right) &=&\left[ \Delta (abc)\left( a+\alpha \right) !\left( a-\alpha
\right) !\left( b+\beta \right) !\left( b-\beta \right) !\left( c+\gamma
\right) !\left( c-\gamma \right) !\right] ^{\frac{1}{2}} \nonumber \\
&&\times \sum\limits_z\frac{\left( -1\right) ^{z+a-b-\gamma }}{\left( 
\begin{array}{c}
z!\left( a+b-c-z\right) !\left( c-b+\alpha +z\right) !\left( b+\beta
-z\right) ! \\ 
\times \left( a-\alpha -z\right) !\left( c-a-\beta +z\right)!
\end{array}
\right) }
\end{eqnarray}
where the summation is over all integers $z$ such that the factorial
arguments are non-negative and 
\begin{displaymath}
\Delta \left( abc\right) =\frac{\left( a+b-c\right) !\left( a-b+c\right)
!\left( -a+b+c\right) !}{\left( a+b+c+1\right) !}.
\end{displaymath}

Let $j$ be a non-negative half integer and let $\left| m^{\prime }\right|
,\left| m\right| \leq j$, then the rotation matrices are defined by 
\begin{eqnarray*}
d_{m^{\prime }m}^j\left( \theta \right) &=&\left( -1\right) ^{j-m^{\prime
}}\left[ \left( j+m^{\prime }\right) !\left( j-m^{\prime }\right) !\left(
j+m\right) !\left( j-m\right) !\right] ^{\frac 12} \\
&&\times \sum_k\left( -1\right) ^k\frac{\left( \cos \frac \theta 2\right)
^{2k+m^{\prime }+m}\left( \sin \frac \theta 2\right) ^{2j-2k-m^{\prime }-m}}{%
k!\left( j-m^{\prime }-k\right) !\left( j-m-k\right) !\left( m^{\prime
}+m+k\right) !}
\end{eqnarray*}
where the summation is over all integers $k$ such that the factorial
arguments are non-negative. The rotation matrices are the matrix elements in
the $2j+1$ dimensional representation of rotations about a fixed
axis.

We have divided the discussion of the relationship between the $3j$ symbol
and the rotation matrix into two cases depending upon $\gamma $. We discuss
the case where $c-\left| \gamma \right| \rightarrow \infty $ in theorem $1$
and the case where $c=\pm \gamma $ in theorem $2$.

\begin{thm}
Fix $b$ and $\beta $ and let $a,c\rightarrow \infty $ in such a way that $%
c-a=\delta $ is fixed and let $c-\left| \gamma \right| \rightarrow \infty $,
then 
\begin{displaymath}
\sqrt{a+b+c+1}\left( 
\begin{array}{ccc}
b & a & c \\ 
\beta & \alpha & \gamma
\end{array}
\right) -(-1)^{b-a-\gamma }d_{\delta ,-\beta }^b\left( \theta \right)
\rightarrow 0
\end{displaymath}
where 
\begin{equation} \label{9}
\cos \theta =\frac {\gamma}{c+\frac{1}{2}}.
\end{equation}
\end{thm}
The theorem is nontrivial as the right hand side does not converge to zero under this asymptotic limit.

\textit{Proof}. Eliminating $a$ and $\alpha $ from (\ref{8}), as $c-a=\delta $ and $\alpha+\beta +\gamma =0$, and then expanding the $\Delta $ symbol gives 
\begin{eqnarray} \label{10}
\left( 
\begin{array}{ccc}
b & a & c \\ 
\beta & \alpha & \gamma
\end{array}
\right) &=&\left( -1\right) ^{b-a-\gamma }\left[ \frac{\left( b+\beta
\right) !\left( b-\beta \right) !\left( b-\delta \right) !\left( b+\delta
\right) !}{2c+b-\delta +1}\right] ^{\frac {1}{2}} \nonumber \\ 
&&\times \sum\limits_z\frac{\left( -1\right) ^z\left[ \Omega _1\Omega
_2\Omega _3\right] ^{\frac{1}{2}}}{z!\left( b-\delta -z\right) !\left( b-\beta
-z\right) !\left( \beta +\delta +z\right) !}
\end{eqnarray}
where 
\begin{eqnarray*}
\Omega _1 &=&\frac{\left( 2c-\delta -b\right) !}{\left( 2c-\delta +b\right) !%
} \\
\Omega _2 &=&\frac{\left( c+\gamma -\delta +\beta \right) !\left( c+\gamma
\right) !}{\left[ \left( c+\gamma -b+\beta +z\right) !\right] ^2} \\
\Omega _3 &=&\frac{\left( c-\gamma -\delta -\beta \right) !\left( c-\gamma
\right) !}{\left[ \left( c-\gamma -\delta -\beta -z\right) !\right] ^2}.
\end{eqnarray*}

As $a$ and $c$ tend to infinity in the manner defined above, it is important
that the range of summation in (\ref{10}) remains finite. If this is not the case then the error generated by the approximations used to reduce the $3j$ symbol to a rotation matrix may not converge to zero. Brussaard and Tolhoek [$4$] did not discuss the effect the asymptotic limit had on the range of summation. The summation variable $z$ is bounded above by 
\begin{displaymath}
\min \left( c-\delta -\gamma -\beta ,b-\delta ,b-\beta \right)
\end{displaymath}
and is bounded below by 
\begin{displaymath}
\max \left( -c+b-\gamma -\beta ,0,-\beta -\delta \right) 
\end{displaymath}
By hypothesis, $c-\left| \gamma \right| \rightarrow \infty $ so that the
first term in each of the above is redundant as $b,\beta $ and $\delta $ are
all fixed. Thus, the range of summation is fixed eventually.

Stirling's approximation \cite{hilb} to $n!$ is 
\begin{displaymath}
n!\sim \sqrt{2\pi }n^{n+\frac{1}{2}}e^{-n}
\end{displaymath}
where $a\sim b$ means $\frac{a}{b}\longrightarrow 1$ as $n$ tends to
infinity. Also, 
\begin{equation} \label{11}
n!\sim \sqrt{2\pi }\left( n+1\right) ^{n+\frac{1}{2}}e^{-\left( n+1\right) }
\end{equation}
and 
\begin{equation} \label{12}
n!\sim \sqrt{2\pi }\left( n+\frac{1}{2}\right) ^{n+\frac{1}{2}}e^{-\left(
n+\frac{1}{2}\right) }.
\end{equation}
Applying (\ref{11}) to the first factor in $\Omega $ and (\ref{12}) to the rest
gives
\begin{equation} \label{13}
\Omega \sim \left( \sin \frac{\theta}{2}\right) ^{2z+\beta +\delta }\left(
\cos \frac{\theta}{2}\right) ^{2b-2z-\beta -\delta }
\end{equation}
where $\theta $ defined by (\ref{9}) and $\Omega =\left[ \Omega _1\Omega
_2\Omega _3\right] ^{\frac 12}.$ The constraint that the factorial arguments
in (\ref{8}) have to be non-negative implies that the above powers are also non-negative as 
\begin{eqnarray*}
2z+\beta +\delta &=&z+(z+\beta +\delta ) \\
2b-2z-\beta -\delta &=&(b-\beta -z)+(b-\delta -z).
\end{eqnarray*}
The right hand side of (\ref{13}) is bounded above by $1$ and so 
\begin{displaymath}
\Omega -\left( \sin \frac {\theta}{2}\right) ^{2z+\beta +\delta }\left( \cos
\frac{\theta}{2}\right) ^{2b-2z-\beta -\delta }\rightarrow 0.
\end{displaymath}

As the range of summation in (\ref{10}) is eventually fixed, the error generated by applying various approximations to the factorials in $\Omega _i,i=1,2,3$, will converge to zero. Thus, 
\begin{displaymath}
\sqrt{a+b+c+1}\left( 
\begin{array}{ccc}
b & a & c \\ 
\beta & \alpha & \gamma
\end{array}
\right) -(-1)^{b-a-\gamma }(-1)^{\beta -b}d_{\beta ,\delta }^b\left( \pi
-\theta \right) \rightarrow 0
\end{displaymath}
and as $d_{\beta ,\delta }^b\left( \pi -\theta \right) =\left( -1\right)
^{b-\beta }d_{\delta ,-\beta }^b\left( \theta \right) $ the result follows. $\Box $

\begin{thm}
Fix $b$ and $\beta $, let $\gamma =\pm c$ and let $%
a,c\longrightarrow \infty $ so that $c-a=\delta $ is fixed, then 
\begin{displaymath}
\sqrt{a+b+c+1}\left( 
\begin{array}{ccc}
b & a & c \\ 
\beta & \alpha & \gamma
\end{array}
\right) -(-1)^\xi \delta _{\alpha ,\mp a}\longrightarrow 0
\end{displaymath}
where 
\begin{displaymath}
\xi =\left\{ 
\begin{array}{ll}
-a+b-c & \mathrm{ for}\;\gamma =c \\ 
2b+\delta -\beta & \mathrm{ for}\;\gamma =-c.
\end{array}
\right.
\end{displaymath}
\end{thm}
\textit{Proof}. Similar to that for theorem $1$. $\Box $

\section{Rotation matrices and $6j$ symbols}

Following Edmonds [$1$], we define the $6j$ symbol as follows. Let $%
a,b,c,d,e $ and $f$ be non-negative half integers such that each of the
triples $\{abc\},\{cde\},\{aef\}$ and $\{bdf\}$ satisfies the triangle
inequality and the sum of the entries in each triple is an integer, then the $6j$ symbol is defined by 
\begin{eqnarray*}
\left\{ 
\begin{array}{lll}
a & b & c \\ 
d & e & f
\end{array}
\right\} &=&\left[ \triangle \left( abc\right) \triangle \left( cde\right)
\triangle \left( aef\right) \triangle \left( bdf\right) \right] ^{\frac 12}
\\
&&\times \sum\limits_z\frac{\left( -1\right) ^z\left( z+1\right) !}{\left( 
\begin{array}{c}
\left( z-a-b-c\right) !\left( z-a-e-f\right) !\left( b+c+e+f-z\right) !\times
\\ 
\left( z-d-e-c\right) !\left( z-b-d-f\right) !\left( a+b+d+e-z\right) !\times
\\ 
\left( a+c+f+d-z\right) !
\end{array}
\right) }
\end{eqnarray*}
where the summation is over all integers $z$ such that the factorial
arguments are non-negative and 
\begin{displaymath}
\triangle \left( abc\right) =\frac{\left( a+b-c\right) !\left( a-b+c\right)
!\left( -a+b+c\right) !}{\left( a+b+c+1\right) !}.
\end{displaymath}
\begin{thm} Fix $f,m,n$ and let $a=\lambda a_0,b=\lambda b_0$ and $%
c=\lambda c_0$ with $\lambda \rightarrow \infty $, then 
\begin{displaymath}
\sqrt{\omega _a\omega _b}\left\{ 
\begin{array}{ccc}
c & b & a \\ 
f & a+n & b+m
\end{array}
\right\} -\left( -1\right) ^\xi d_{nm}^f\left( \phi \right) \rightarrow 0
\end{displaymath}
where 
\begin{displaymath}
\omega _a=2a+n+f+1,\;\omega _b=2b+m+f+1,\;\xi =a+b+c+f+m
\end{displaymath}
and 
\begin{displaymath}
\cos \phi =\frac{\left( a+n+\frac{1}{2}\right) ^2+\left( b+m+\frac{1}{2}\right)
^2-\left( c+\frac{1}{2}\right) ^2}{2\left( a+n+\frac{1}{2}\right) \left(
b+m+\frac{1}{2}\right) }.
\end{displaymath}
\end{thm}

The scaling used in theorem $3$ is clearly different from that used in
theorem $1$. The reason for this is that as $\lambda $ tends to infinity, $%
\phi $ is preserved, whereas, if we had scaled $a_0,b_0$ and $c_0$ as we did
in theorem $1$ then $\phi $ would not have been preserved.
Theorem $3$ is non-trivial as $d_{nm}^f\left( \phi \right) $ does not
converge to zero under this asymptotic limit.

\textit{Proof}. Changing the summation variable from $z$ to $t$, where $z=a+b+c+m+n+t$, and
then expanding the $\triangle $ symbols gives, 
\begin{eqnarray*}
\sqrt{\omega _a\omega _b}\left\{ 
\begin{array}{ccc}
c & b & a \\ 
f & a+n & b+m
\end{array}
\right\} &=&\left( -1\right) ^\nu \left[ \left( f+m\right) !\left(
f-m\right) !\left( f+n\right) !\left( f-n\right) !\right] ^{\frac 12} \\
&&\times \sum\limits_t\frac{\left( -1\right) ^t\left[
\prod\limits_{i=1}^6\Phi _i\right] ^{\frac{1}{2}}}{t!\left( f-m-t\right)
!\left( f-n-t\right) !\left( m+n+t\right) !}
\end{eqnarray*}
where $\nu =a+b+c+m+n$ and 
\begin{eqnarray*}
\Phi _1 &=&\frac{\left( 2a+n-f\right) !}{\left( 2a+n+f\right) !} \\
\Phi _2 &=&\frac{\left( 2b+m-f\right) !}{\left( 2b+m+f\right) !} \\
\Phi _3 &=&\frac{\left( a+b-c\right) !\left( a+b-c+m+n\right) !}{\left[
\left( a+b-c-t\right) !\right] ^2} \\
\Phi _4 &=&\frac{\left( a-b+c\right) !\left( a-b+c+n-m\right) !}{\left[
\left( a-b+c+n-f+t\right) !\right] ^2} \\
\Phi _5 &=&\frac{\left( -a+b+c\right) !\left( -a+b+c+m-n\right) !}{\left[
\left( -a+b+c+m-f+t\right) !\right] ^2} \\
\Phi _6 &=&\frac{\left[ \left( a+b+c+m+n+1+t\right) !\right] ^2}{\left(
a+b+c+1\right) !\left( a+b+c+m+n+1\right) !}.
\end{eqnarray*}
As in the proof of theorem $1$, the constraints on $t$ ensure that as $a,b$
and $c$ tend to infinity in the manner defined above, the range of summation
remains finite so that the error generated by applying various
approximations to the factorials in $\Phi _i,i=1,2,...,6,$ will converge to
zero. By applying (\ref{11}) to $\Phi _1$ and $\Phi _2$ and (\ref{12}) to the rest and following the arguments used in the proof of theorem $1$ the result
follows. $\Box $

\section{Discussion}

In the first part of this section we discuss why $\phi $ and $\theta $ are
defined by (\ref{6}) and (\ref{7}) rather than by (\ref{2}) and (\ref{4}). In the last part of this section we discuss the relationship between theorems $1$ and $3$ which entails using theorem $1$ to arrive at an expression of a similar type to that in theorem $3$.

In the introduction we noted that there was a discrepancy between the
geometrical and algebraic parts of the asymptotics of the $6j$ symbol.
Edmonds defined $\phi $ and $\theta $ by (\ref{2}) and (\ref{4}) respectively, whereas the geometrical interpretation of the asymptotics, given by Ponzano and Regge, leads to $\phi $ and $\theta $ being defined by (\ref{6}) and (\ref{7}). Of course, (\ref{2}) is asymptotic to (\ref{6}), and (\ref{4}) is asymptotic to (\ref{7}), because, if we substitute the binomial expansion of $\sqrt{j(j+1)}$ into each of the terms in (\ref{2}) and (\ref{4}) then (\ref{2}) will converge to (\ref{6}), and (\ref{4}) will converge to (\ref{7}), but, the geometry which leads to $\phi $ and $\theta $ being defined as in (\ref{2}) and (\ref{4}), when applied to a certain asymptotic formula, leads to a worse approximation to the $6j$ symbol. In [$4$], Ponzano and Regge give an asymptotic formula for the $6j$ symbol in which all the entries all large. They label the edges of the tetrahedron associated to this asymptotic $6j$ symbol by $j+\frac 12$ where $j$ is an entry in the $6j$ symbol. Suppose we now relabel the edges of the this tetrahedron by $\sqrt{j(j+1)}$ and apply the relevant modifications to the Ponzano-Regge asymptotic formula. Then as can be seen in the table below, this modified asymptotic formula provides a worse approximation to the $6j$ symbol than Ponzano and Regge's original formula. Thus the labelling of the edges of the associated tetrahedron by $j+\frac 12$ is critical in the Ponzano-Regge asymptotic formula and as $(1)$ and $(3)$ should give rise to an asymptotic formula of the Ponzano-Regge type under some further asymptotic limit, we shall defined $\phi $ and $\theta $ as in (\ref{6}) and (\ref{7}). Also, by defining $\phi $ and $\theta $ by (\ref{6}) and (\ref{7}), all the formulae are simpler, which can be seen in the proofs of theorems $1$ and $3$. 

\begin{center}
\begin{tabular}{|c|c|c|c|c|c|c|c|c|}
\hline
$a$ & $b$ & $c$ & $d$ & $e$ & $f$ & $6j$ & $j+\frac 12$ & $\sqrt{j(j+1)}$ \\ 
\hline
$1$ & $1$ & $1$ & $1$ & $1$ & $1$ & $0.16666667$ & $0.16682679$ & $%
-0.07825688$ \\ \hline
$\frac {7}{2}$ & $7$ & $\frac{9}{2}$ & $\frac{17}{2}$ & $5$ & $\frac{5}{2}$ & $
-0.04178554$ & $-0.04152025$ & $-0.02937347$ \\ \hline
$\frac{17}{2}$ & $\frac{15}{2}$ & $10$ & $\frac{15}{2}$ & $\frac{15}{2}$ & $4$ & $
0.01649429$ & $0.01642250$ & $0.01293466$ \\ \hline
$\frac{13}{2}$ & $8$ & $\frac{9}{2}$ & $\frac{13}{2}$ & $6$ & $\frac{15}{2}$ & $
0.02551804$ & $0.02550576$ & $0.02284036$ \\ \hline
$5$ & $8$ & $12$ & $9$ & $7$ & $6$ & $-0.02244118$ & $-0.02242208$ & $
-0.01944464$ \\ \hline
$9$ & $9$ & $9$ & $9$ & $9$ & $9$ & $-0.01565006$ & $-0.01564009$ & $
-0.01485678$ \\ \hline
$7$ & $8$ & $9$ & $6$ & $9$ & $5$ & $-0.01370361$ & $-0.01321203$ & $
-0.00872052$ \\ \hline
$7$ & $8$ & $9$ & $6$ & $9$ & $14$ & $0.00166374$ & $0.00166376$ & $
0.01294491$ \\ \hline
$13$ & $15$ & $24$ & $\frac{29}{2}$ & $\frac{33}{2}$ & $\frac{19}{2}$ & $
-0.00852912$ & $-0.00856189$ & $-0.00892974$ \\ \hline
$13$ & $15$ & $24$ & $\frac{29}{2}$ & $\frac{33}{2}$ & $\frac{29}{2}$ & $
-0.00562296$ & $-0.00543266$ & $-0.00471280$ \\ \hline
\end{tabular}
\end{center}

The column labelled $6j$ contains the exact value of the $6j$ symbol 
\begin{displaymath}
\left\{ 
\begin{array}{lll}
a & b & c \\ 
d & e & f
\end{array}
\right\}
\end{displaymath}
as defined in section $3$. The next column contains the Ponzano-Regge
approximation to the $6j$ symbol in which all the entries are large [$4$], 
\begin{displaymath}
\left\{ 
\begin{array}{lll}
a & b & c \\ 
d & e & f
\end{array}
\right\} \simeq \frac{1}{\sqrt{12\pi V}}\cos \left(
\sum\limits_{h<k}j_{hk}\theta _{hk}+\frac{\pi}{ 4}\right)
\end{displaymath}
where 
\begin{eqnarray*}
j_{12} &=&a+\frac{1}{2},\;j_{13}=b+\frac{1}{2},\;j_{14}=c+\frac{1}{2}\\j_{23} &=&d+\frac{1}{2},\;j_{24}=e+\frac{1}{2}\textrm{ and }j_{34}=f+\frac{1}{2}
\end{eqnarray*}
are the edge lengths of the associated tetrahedron and $\theta _{hk}$ is the
exterior dihedral angle between the two planes on the associated tetrahedron
which meet in the edge $j_{hk}$, and $V$ is the volume of the associated
tetrahedron. The last column contains the modified Ponzano-Regge formula in
which $\sqrt{j(j+1)}$ has replaced $j+\frac{1}{2}$ everywhere, including in the
calculation of the volume $V$ and the exterior dihedral angles $\theta _{hk}$.

Superficially, theorems $1$ and $3$ appear to be similar as they both
involve the reduction of the primary object to a rotation matrix. It is thus
natural to ask if it is possible to arrive at an expression of a similar
type to that in theorem $3$ by using theorem $1$. This is indeed possible,
but it is important to point out that the asymptotic limit in this result
will be the limit used in theorem $1$ which is not the same as the scaling
used in theorem $3$.

The following arguments are not intended to prove the link between theorems $%
1$ and $3$ but are to be regarded as a possible justification for there
being a link.

To discuss the existence of such a link we shall use two limits, both of
which are of the same type used in theorem $1$. The first limit reduces the
relevant $6j$ symbol to a $3j$ symbol [$4$], and the second limit when
applied to this $3j$ symbol reduces it, by theorem $1$, to a rotation
matrix. The overall result is an expression of the same type as in theorem $%
3 $ but with a different scaling of the nonnegative half integers.

In [$4$], Ponzano and Regge discuss how the $6j$ symbol 
\begin{displaymath}
\left\{ 
\begin{array}{ccc}
a & b & c \\ 
d+R & e+R & f+R
\end{array}
\right\}
\end{displaymath}
reduces to the $3j$ symbol 
\begin{displaymath}
\left( 
\begin{array}{lll}
a & b & c \\ 
\alpha & \beta & \gamma
\end{array}
\right)
\end{displaymath}
as $R\rightarrow \infty $, where $\alpha =e-f,\beta =f-d$ and $\gamma =d-e$.
If we expand the particular factorials in the above $6j$ symbol suggested by
Ponzano and Regge, using (\ref{11}) and (\ref{12}), and then follow the proof of theorem $1$, this shows that, as $R\rightarrow
\infty $, 
\begin{equation} \label{14}
\sqrt{2R+1}\left\{ 
\begin{array}{ccc}
a & b & c \\ 
d+R & e+R & f+R
\end{array}
\right\} -(-1)^\chi \left( 
\begin{array}{lll}
a & b & c \\ 
\alpha & \beta & \gamma
\end{array}
\right) \rightarrow 0
\end{equation}
where $\chi =a+b+c+2(d+e+f)$. This represents a rigorous statement of
Ponzano and Regge's work.

\begin{figure}
\hspace{3cm} \epsfxsize=8cm \epsfbox{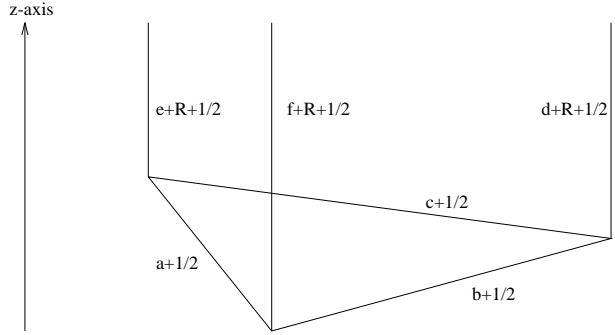}
\caption{\label{fig4} Associated tetrahedron}
\end{figure}

The geometrical interpretation of the above result is discussed by Ponzano
and Regge in [$4$]. As $R\rightarrow \infty $ in the above $6j$ symbol,
three of the vertices of the associated tetrahedron remain fixed and the
remaining one moves further and further away from the others. This causes
the edges labelled by $d,e$ and $f$ to become parallel to each other and it
is these edges which define a $z$ axis, with $\alpha ,\beta $ and $\gamma $
being the projections of the edges labelled by $a,b$ and $c$ onto this axis (see figure \ref{fig4}). The triangle associated to the above $3j$ symbol is the unscaled triangle in the associated tetrahedron formed by the three fixed vertices.

Now, replace $a$ by $b$, $b$ by $b+m$, $c$ by $f$, $d$ by $a+n$, $e$ by $a$
and $f$ by $c$ in (\ref{14}) and apply the symmetries 
\begin{displaymath}
\left\{ 
\begin{array}{lll}
a & b & c \\ 
d & e & f
\end{array}
\right\} =\left\{ 
\begin{array}{lll}
f & a & e \\ 
c & d & b
\end{array}
\right\}
\end{displaymath}
and 
\begin{displaymath}
\left( 
\begin{array}{lll}
a & b & c \\ 
\alpha & \beta & \gamma
\end{array}
\right) =\left( 
\begin{array}{lll}
c & a & b \\ 
\gamma & \alpha & \beta
\end{array}
\right) ,
\end{displaymath}
which gives 
\begin{equation} \label{15}
\sqrt{\omega _1}\left\{ 
\begin{array}{ccc}
c+R & b & a+R \\ 
f & a+n+R & b+m
\end{array}
\right\} -(-1)^\xi \left( 
\begin{array}{ccc}
f & b & b+m \\ 
n & a-c & c-a-n
\end{array}
\right) \rightarrow 0
\end{equation}
where 
\begin{displaymath}
\omega _1=2R+1\;\mathrm{ and }\;\xi =2b+m+f+2n+2(c+R).
\end{displaymath}

We now scale $b$ by adding to it a real number $R^{\prime }$ which is
tending to infinity. Thus, by theorem $1$, 
\begin{equation} \label{16}
\sqrt{\omega _2}\left( 
\begin{array}{ccc}
f & b+R^{\prime } & b+m+R^{\prime } \\ 
n & a-c & c-a-n
\end{array}
\right) -\left( -1\right) ^\mu d_{nm}^f\left( \phi \right) \rightarrow 0
\end{equation}
where 
\begin{displaymath}
\omega _2=2(b+R^{\prime })+m+f+1\;\mathrm{ and }\;\mu =(a+R)-(b+R^{\prime
})-(c+R)+2f
\end{displaymath}
and $\phi $ is the angle between the edges labelled by $a+n+R$ and $%
b+m+R^{\prime }$ in the tetrahedron associated to the above $6j$ symbol.
Equations (\ref{15}) and (\ref{16}) give 
\begin{displaymath}
\sqrt{\omega _1\omega _2}\left\{ 
\begin{array}{ccc}
c+R & b+R^{\prime } & a+R \\ 
f & a+n+R & b+m+R^{\prime }
\end{array}
\right\} -\left( -1\right) ^\tau d_{nm}^f\left( \phi \right) \rightarrow 0
\end{displaymath}
where 
\begin{displaymath}
\tau =(a+R)+(b+R^{\prime })+(c+R)+f+m
\end{displaymath}
and 
\begin{displaymath}
\cos \phi =\frac{\left( a+n+R+\frac{1}{2}\right) ^2+\left( b+m+R^{\prime
}+\frac{1}{2}\right) ^2-\left( c+R+\frac{1}{2}\right) ^2}{2\left( a+n+R+\frac{1}{2}
\right) \left( b+m+R^{\prime }+\frac{1}{2}\right) }.
\end{displaymath}
This result is of the same type as Theorem $3$ but with two scalings.

In the introduction we had figure $3$ which represented the different
asymptotic limits of the $6j$ symbol in terms of the tetrahedron associated
to that $6j$ symbol. The above result is represented by the arrow from $4$
to $1+1+2$ via $1+3$. The arrow from $4$ to $1+3$ is Ponzano and Regge's
reduction of a $6j$ symbol to a $3j$ symbol in which one vertex of the
associated tetrahedron is taken to infinity while the remaining three are
fixed, and the arrow from $1+3$ to $1+1+2$ is where one of the three
previously fixed vertices is taken to infinity. Hence the present paper has
completed the triangle with vertices $4,1+3$ and $1+1+2$.

\section*{Acknowledgments}

The author would like to thank John Barrett for several helpful discussions. Figure 3 is produced here courtesy of John Barrett. The author is supported by an EPSRC research studentship.

\bibliography{jmp32}

\begin{thebibliography}{1}

\bibitem{edm55}
A~R Edmonds.
\newblock {\em Angular momentum in quantum mechanics}.
\newblock CERN 55-26 Geneve, 1955.

\bibitem{racah51}
G~Racah.
\newblock {\em Group theory and spectroscopy}, page~74.
\newblock Inst. Adv. Study Princeton, 1951.

\bibitem{bruss}
P~J Brussaard and H~A Tolhoek.
\newblock {\em Physica}, pages 955--971, 1957.

\bibitem{pr}
G~Ponzano and T~Regge.
\newblock Semiclassical limits of racah coefficients.
\newblock In F~Block, editor, {\em Spectroscopic and group theoretical methods
  in physics}, pages 1--58. North-Holland Amsterdam, 1968.

\bibitem{jwbc97}
J~W Barrett and L~Crane.
\newblock An algebraic interpretation of the wheeler-dewitt equation.
\newblock {\em Class. Quantum Grav}, pages 2113--2121, 1997.

\bibitem{jwbfig}
J~W Barrett.
\newblock Private communication, 1997.

\bibitem{hilb}
R~Courant and D~Hilbert.
\newblock {\em Methods of mathematical physics}.
\newblock New York, 1966.

\end{thebibliography}

\end{document}